\documentclass{Interspeech}

  \interspeechcameraready

\title{A Data-Driven Diffusion-based Approach for Audio Deepfake Explanations}

\author[affiliation={1,2}]{Petr}{Grinberg}
\author[affiliation={2}]{Ankur}{Kumar}
\author[affiliation={2}]{Surya}{Koppisetti}
\author[affiliation={2}]{Gaurav}{Bharaj}

\affiliation{}{EPFL}{Switzerland}
\affiliation{}{Reality Defender Inc.}{USA}
\email{petr.grinberg@epfl.ch, \{ankur, surya, gaurav\}@realitydefender.ai}

\usepackage{comment}

\usepackage{multirow}

\newcommand{\astfootnote}[1]{%
  \begingroup%
  \renewcommand{\thefootnote}{\fnsymbol{footnote}}%
  \footnote[1]{#1}%
  \endgroup%
}

\begin{document}

\maketitle

\begin{abstract} 

    Evaluating explainability techniques, such as SHAP and LRP, in the context of audio deepfake detection is challenging due to lack of clear ground truth annotations. In the cases when we are able to obtain the ground truth, we find that these methods struggle to provide accurate explanations.
    In this work\astfootnote{This work was done during an internship at Reality Defender Inc.}, we propose a novel data-driven approach to identify artifact regions in deepfake audio. We consider paired real and vocoded audio, and use the difference in time-frequency representation as the \emph{ground-truth explanation}. The difference signal then serves as a supervision to train a diffusion model to expose the deepfake artifacts in a given vocoded audio. Experimental results on the VocV4 and LibriSeVoc datasets demonstrate that our method outperforms traditional explainability techniques, both qualitatively and quantitatively. 
    
\end{abstract}

\section{Introduction}\label{sec:introduction}
Explainable AI (XAI) methods such as GradCAM~\cite{selvaraju2017grad}, SHAP~\cite{ge2022explaining, ge2022explainable} and LRP~\cite{achtibat2024attnlrp} have been used in the past to attribute model decisions in a variety of audio tasks. 
For some tasks, such as sound classification, we can listen to audio segments corresponding to the highlighted regions to check if the sound matches the label~\cite{paissan2024listenable}. However, for the audio deepfake detection (ADD) task, it is often unclear what portions of an utterance correspond to manipulation, making it difficult for humans to interpret and evaluate the attributions \cite{zhang2024common}. 
Recent work \cite{grinberg2025does} shows that existing post-hoc XAI tools do not provide consistent explanations in the time domain, which further necessitates careful evaluation of these methods.

Vocoders are integral component in many text-to-speech and voice conversion systems~\cite{wang2023spoofed}. Several vocoded datasets exist, such as VocV4~\cite{wang2023spoofed} and LibriSeVoc~\cite{sun2023ai}, which contain parallel real and fake audios with the same  speaker and time-synchronized spoken content. The fake audio is synthesized by applying a vocoder on an intermediate representation, such as spectrogram, derived from the corresponding real audio. Therefore, the only difference between a real audio and its vocoded version is the introduction of artifacts from the vocoder. We can utilize this difference to create \emph{ground truth explanations}, which enables the evaluation of above explainability methods. Also, as we show later in Section \ref{sec:experiments}, the classical XAI tools struggle to provide accurate explanations, which indicates the need for methods that better align with the ground truth.

In the field of vision deepfakes, Zhang et al.~\cite{zhang2024common} proposed a trainable visual question answering framework to provide textual explanations for unnatural deepfake features, like blurry hairline, that align with commonsense reasoning. Trainable XAI methods have also been used to provide faithful interpretations for the audio event classification task~\cite{paissan2024listenable, mancini2024lmac}. 
Inspired by these works, we propose a data-driven XAI method that can be trained using the difference between real and vocoded utterances as the supervisory signal to account for the specificity of the deepfake detection task.
Our main contributions are:

\begin{enumerate}
    \item We propose a methodology to generate annotations for ADD XAI using  parallel samples in a vocoded dataset, i.e., pairs of fake and real utterances with the same speaker and time-synchronized content.
    \item We show that a classifier-agnostic diffusion model can be trained on spoof spectrogram and the corresponding difference signal as the input-output pair to predict heatmaps highlighting the most suspicious regions in the time-frequence domain.
    Also, we show that post-hoc classifier-specific attributions can be obtained by conditioning the diffusion model on intermediate features from the same classifier.
    \item We extend the classical XAI tools to output time-frequency attribution and show that the proposed diffusion model produces attributions that are more faithful to the model and are better aligned with the ground-truth annotations.
\end{enumerate}

\section{Related Work}

While several works investigated the explainability of ADD systems~\cite{ge2022explaining, ge2022explainable,
salvi2023towards, 
li2024audio},
only few~\cite{liu2024neural, li2024interpretable, grinberg2025does} have considered state-of-the-art models, which directly operate on raw waveforms~\cite{li2024audio}. 
These works are further limited to time-domain explanations only, which allow for a partial understanding of the model behavior, as artifacts can be found in specific frequency sub-bands~\cite{tak2020explainability}. In the paper, we provide explanations in the joint time-frequency domain for a state-of-the-art deepfake classifier~\cite{tak2022automatic}, which operates on raw waveforms.

Previous works~\cite{frank2021wavefake, wang2023spoofed, sun2023ai, wang2024can}
have either used vocoded data as a training source for the ADD systems themselves~\cite{wang2023spoofed, sun2023ai, wang2024can} or have resorted to exploratory studies on the statistical differences between real and fake samples, such as average pitch or energy across frequency bins~\cite{frank2021wavefake}.
In contrast, we consider the difference signal in parallel pairs as the ground-truth explanation and use it as a supervisory signal to train a diffusion model for predicting the spoof artifact regions.

Neural network-based XAI was explored in \cite{paissan2024listenable, mancini2024lmac} using self-supervised losses, whereas we train a diffusion model in a supervised manner using the ground-truth explanations.
Diffusion-based models have been used previously for synthetic image detection~\cite{cazenavette2024fakeinversion, sun2024diffusionfake} and speaker verification~\cite{kim2024diff, bai2024diffusion}. However, these works focus on improving detection capabilities,
whereas we consider diffusion for a trainable XAI technique. 

\section{Methodology}\label{sec:methodology}

\subsection{Classical explainability methods}\label{sec:rawwrapper}

State-of-the-art ADD models today operate on raw waveform inputs~\cite{li2024audio}. To enable explanations in the time-frequency domain with classical XAI methods, we propose the RawWrapper technique, which prepends Inverse Short-Time Fourier Transform (ISTFT) as a convolutional layer at the start of the ADD model.
For an ADD model $f$ operating on raw waveform $\mathbf{x}$, we have:
\begin{align}
    f(\mathbf{x}) = f(\text{ISTFT}(\text{STFT}(\mathbf{x}))) = f(\text{ISTFT}(\mathbf{M}, \mathbf{P}))
\end{align}
where $\mathbf{M}$ and $\mathbf{P}$ are the magnitude and phase spectrograms obtained from $\mathbf{x}$ after applying STFT. By defining $\text{RawWrapper}_{f}(\mathbf{M}, \mathbf{P}) = f(\text{ISTFT}(\mathbf{M}, \mathbf{P}))$, we get a wrapped model that matches behavior of the ADD system but operates on input in the time-frequency domain. Simultaneously, we are able to apply the classical XAI techniques, such as SHAP and LRP, directly in the time-frequency domain as $\mathbf{M}$ becomes the part of the computational graph, which allows propagating relevancy from the output to the input. However, other methods like Grad-CAM~\cite{selvaraju2017grad} and transformer relevancy~\cite{grinberg2025does} still cannot be applied due to their dependency on spatial relationship between intermediate features and the input $\mathbf{M}$, which is not clear. Therefore, we do not consider such methods in this work.
We note that the RawWrapper relies on the exact replication of the $f$ decision-making process, which can be achieved by properly choosing STFT parameters to facilitate perfect reconstruction of a signal via ISTFT. In practice, we operate on the log-magnitude spectrogram for better visualization and value range and take its exponential before ISTFT.

\subsection{Dataset for supervised XAI}\label{sec:dataset_annotation}

\begin{figure}[!tp]
    \centering
    \includegraphics[width=\linewidth]{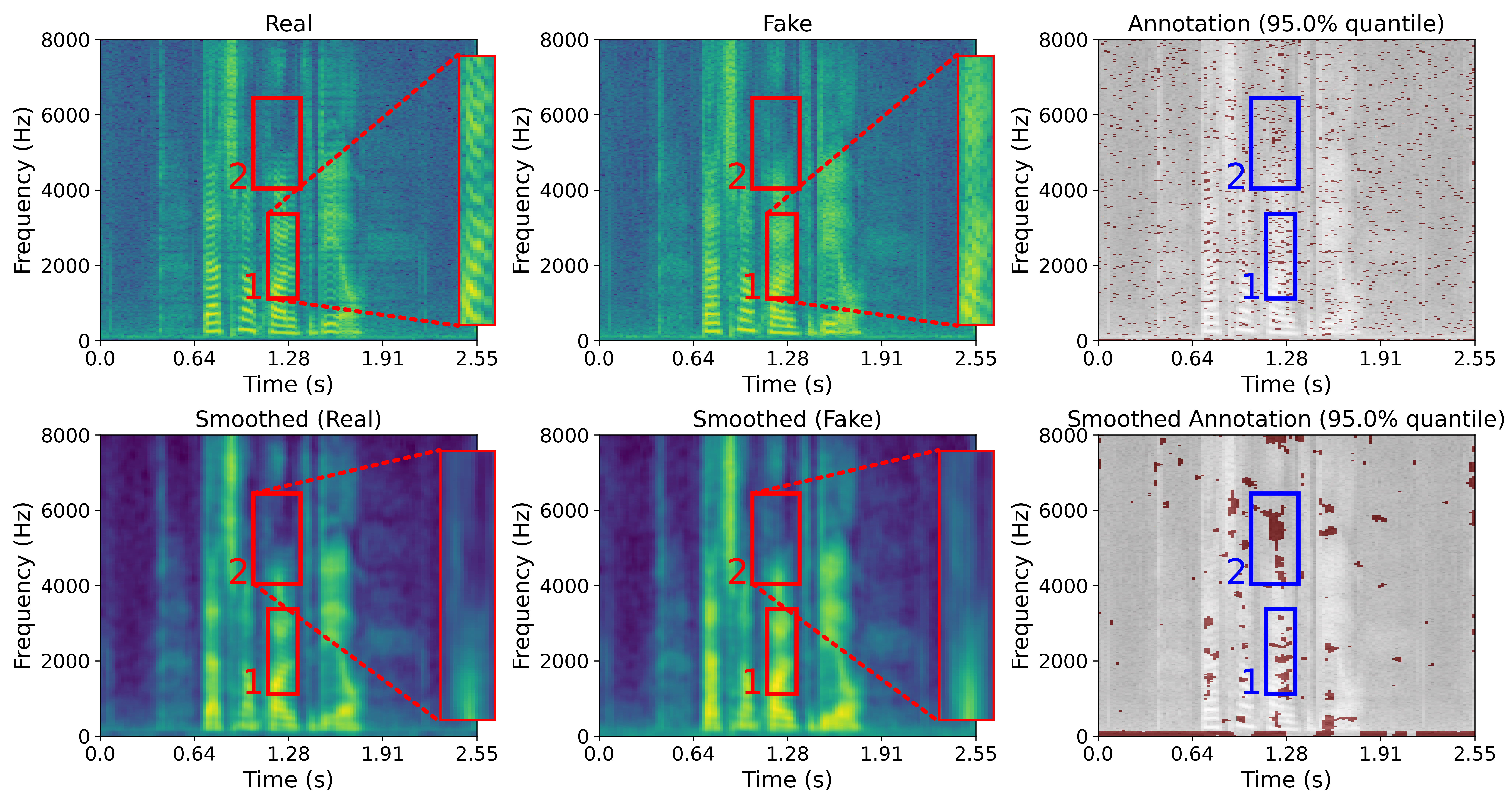}
    \caption{Original (top) and smoothed (bottom) versions of bona fide and spoof spectrograms for a parallel pair in VocV4~\cite{wang2023spoofed},  with corresponding segmentation masks (right). Rectangles show manipulated regions: (1) missed harmonics, (2) energy loss.}
    \label{fig:dataset_annotation}
\end{figure}

If genuine and spoofed speech are perfectly aligned in speaker and content attributes, as in {\em parallel} data pairs, the only distinction 
between the audio frames is the spoofing artifact.
By subtracting bona fide spectrogram $\mathbf{M}_{b}$ from a parallel spoof spectrogram $\mathbf{M}_{s}$, we obtain a heatmap that showcases the most manipulated regions. However, this heatmap tends to be noisy, whereas we are interested in identifying dense regions only since they are more likely to be recognized by humans. Therefore, before computing the difference, we apply smoothing with a 2-D Gaussian kernel $G$ of size $(3, 11)$ and variance of $(3, 5)$ for the time and frequency dimensions, respectively. We also normalize the smoothed difference since the low frequency bins contain more energy (larger magnitude) than the high frequency bins.
The final heatmap is obtained by binarizing the normalized values as follows
\begin{align}
    \text{Mask} = \left(\frac{|G(M_s) - G(M_b)|}{|G(M_b)|}\right) > \tau
    \label{eq:mask}
\end{align}
where $\tau$ is the threshold corresponding to the $95\%$ quantile of the normalized values.
Figure \ref{fig:dataset_annotation} shows the mask obtained using a parallel real and fake sample in the VocV4~\cite{wang2023spoofed} dataset. The top right mask is obtained without applying any smoothing, whereas the mask on the bottom right is acquired when $G$ is applied. 
We note that the regions in the latter case are more dense and meaningful compared to the case when smoothing is not applied. We believe that the above procedure is able to retain most of the significant regions where vocoder artifacts may be present, as well as highlight them faithfully. Therefore, we also call the computed mask as \emph{ground truth explanations}.

\subsection{Diffusion-based XAI}\label{sec:diffusion_xai}
\begin{figure}[t]
    \centering
    \includegraphics[width=0.8\linewidth]{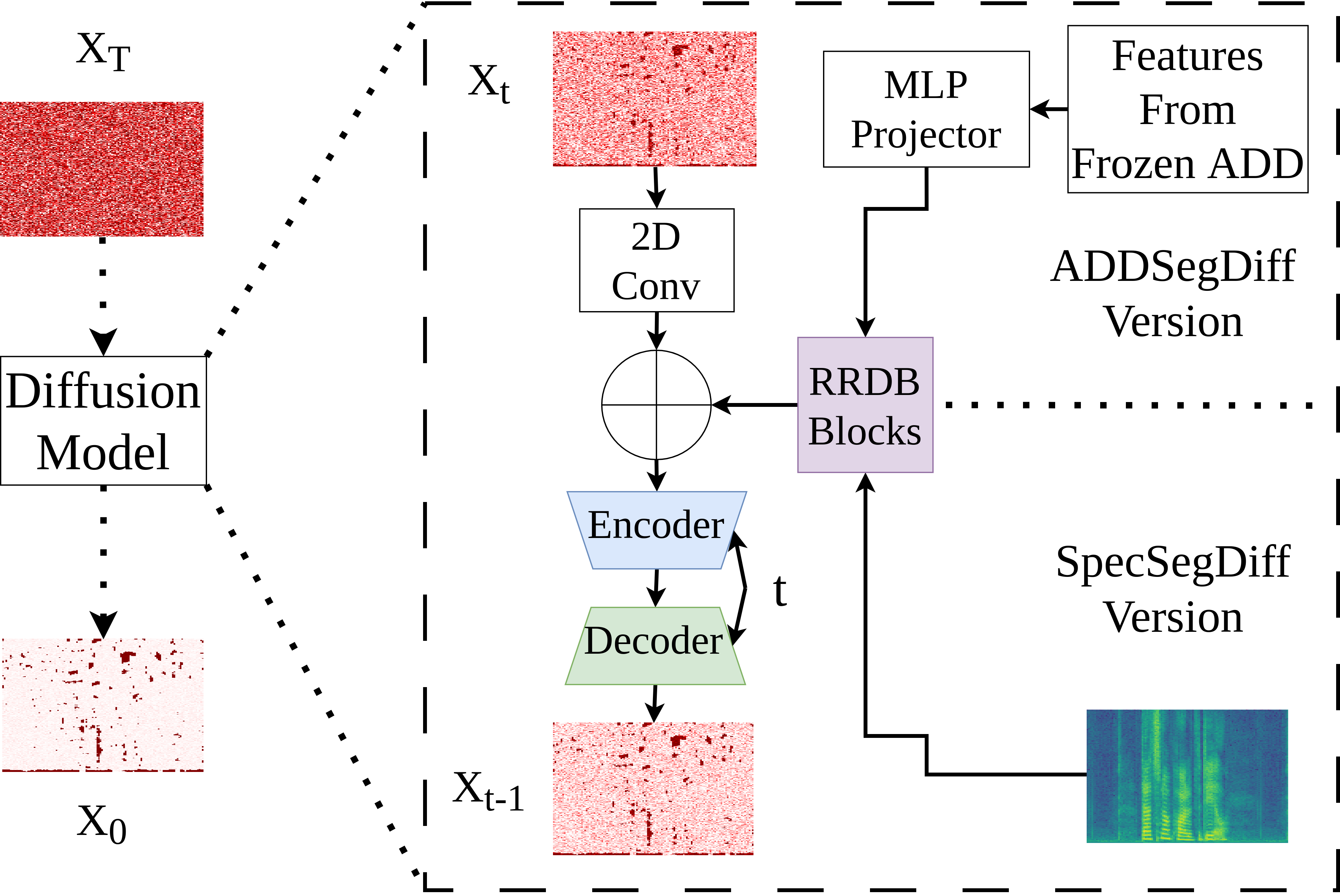}
    \caption{Proposed diffusion model framework to predict artifact regions in spoof audio. For conditioning, we use the spectrogram in SpecSegDiff, whereas, in ADDSegDiff, the intermediate features from a pretrained frozen ADD model are used.}
    \label{fig:segdiff}
\end{figure}

We train a diffusion model to predict spoof artifact regions using ground truth explanations from Section \ref{sec:dataset_annotation} as the target. This task is similar to the standard segmentation task in computer vision. Therefore, we use the SegDiff~\cite{amit2021segdiff} diffusion model\footnote{we base our code on the official implementation \url{https://github.com/tomeramit/SegDiff}} as it has shown high segmentation quality, surpassing non-diffusion approaches on a variety of datasets~\cite{rottensteiner2014isprs, cordts2016cityscapes, kumar2017dataset}, including those with sparse masks as in our annotated datasets.
Figure \ref{fig:segdiff} shows the overall framework. The input to the diffusion model is random noise, which is iteratively denoised to generate the output heatmap based on some conditioning. The diffusion model follows the standard UNet-like encoder-decoder architecture
with 2 residual blocks per level. We refer the readers to \cite{amit2021segdiff} for more details. We experiment with two different ways to condition the generation. First, we use the spectrogram $\mathbf{M}_s$ from spoof audio that is agnostic to the ADD model. We call this setup as SpecSegDiff. This case resembles the scenario when we manually try to identify the artifact regions, regardless of how the ADD model makes its decision. Different from SpecSegDiff, we also try to condition using the intermediate features from the ADD model instead of the spectrogram. If we get similar outputs compared to SpecSegDiff and ground-truth annotations, then we can comment that the features from the ADD model actually utilize the annotated artifacts to make the real vs fake decision.
We call this variation as ADDSegDiff. SpecSegDiff and ADDSegDiff use $12$ and $1$ RRDB~\cite{wang2018esrgan} blocks to preprocess the conditioning, respectively.

\subsection{Evaluation criteria}\label{sec:evaluation_criteria}

We use qualitative and quantitative measures to evaluate the proposed method. 
The former involves manually inspecting the generated heatmap from different methods and comparing them against the binary ground truth obtained in Section \ref{sec:dataset_annotation}. The quantitative measures assess the alignment between predicted heatmaps and the ground truth annotations using standard metrics from the image segmentation task ~\cite{amit2021segdiff, jadon2020survey}: Generalized Dice (GDice), $F_1$-score, IoU,  Boundary $F_1$ (FBound), and Structural Similarity Index Measure (SSIM). Since the diffusion generates attributions only for the spoof class, the evaluation is conducted on the fake subset of the data.

We also compute classic model faithfulness metrics from \cite{paissan2024listenable}, i.e., Average Increase (AI $\uparrow$), Average Drop (AD $\downarrow$), Average Gain (AG $\uparrow$), and Input Fidelity (Fid-In $\uparrow$) to evaluate the change in model's confidence when the input is masked with a heatmap $\mathbf{H}$.
However, since $\mathbf{H}$ is sparse, the resulting masked audio, $\text{ISTFT}(\mathbf{H} \cdot \mathbf{M}_s, \mathbf{P}_s)$, suffers severe distortion that impacts the decision of the ADD model.
To circumvent this issue, we replace the masked regions in magnitude and phase spectrograms with their real counterparts. Therefore, the magnitude and phase inputs to ISTFT becomes $\mathbf{H}\cdot \mathbf{M}_s + (\mathbf{1}-\mathbf{H})\cdot \mathbf{M}_b$ and $\mathbf{H}\cdot \mathbf{P}_s + (\mathbf{1}-\mathbf{H})\cdot \mathbf{P}_b$ respectively.
If a particular region is not important for the model decision, its replacement with a bona fide counterpart should not significantly affect the prediction and vice versa. This means that the drop in the model score should happen only if $\mathbf{H}$ is indeed faithful and highlights the spoof artifact regions.

\section{Experiments}\label{sec:experiments}

\subsection{Experimental setup}

We use VocV4 vocoded dataset to generate the supervised XAI dataset, which is then used to train SpecSegDiff and ADDSegDiff models, as well as compare them against classical XAI methods. Each real utterance in the dataset is re-synthesized using four different vocoders applied on the genuine mel spectrograms. To verify that our methodology can be applied on different datasets, we also use LibriSeVoc~\cite{sun2023ai}, downsampled to \SI{16}{\kilo\hertz}, as the second data source. In contrast to VocV4 that uses generative adversarial network (GAN), flow-based, and signal processing-based vocoders, this dataset consists of six different vocoders created using GAN, diffusion, and autoregressive approaches. In LibriSeVoc, fake audio are created using resynthesis technique but with trimmed silences at the edges, which creates misalignment between real and fake audio pair. We resolve this by applying dynamic time warping to find border points and shift spoof and real utterances in the time domain to get aligned audios.

We train SpecSegDiff until convergence using Adam optimizer with constant learning rate of $10^{-4}$ and weight decay of $10^{-4}$ for around 100K steps. We use global batch size of 24 in all these experiments. For ADDSegDiff experiments, we use Wav2Vec2-AASIST~\cite{tak2022automatic}, a popular ADD model with state-of-the-art performance~\cite{li2024audio},
to extract the conditioning features. The training of the ADD model follows the original work except for the difference in the training dataset, which is VocV4 in this work. The ADD model has an equal error rate (EER) of $1.17\%$, $1.40\%$, $2.75\%$, and $5.04\%$ on VocV4 (development set), LibriSeVoc~\cite{sun2023ai}, ASVspoof 2019~\cite{wang2020asvspoof}, and In-The-Wild~\cite{muller2022does} datasets, respectively. We take the outputs of $\{0, 4, 9, 14, 19, 23\}$-th transformer layers from the frozen ADD model as the conditioning features for ADDSegDiff. We project the features from 1024 to 320 dimensions using 2-layer multilayer perceptron (MLP) before passing them to ADDSegDiff. We train ADDSegDiff for around 100K steps with a global batch size of 48. For all the experiments, we apply RawBoost~\cite{tak2022rawboost} data augmentation with 50\% chance to improve the diffusion model's robustness to clean and noisy utterances.
To ensure that augmented audio still differs only because of vocoder artifacts, fake and its parallel real audio are augmented with the same random noise-vector and hyperparameters.
During inference, each diffusion model predicts 32 masks using different random noise samples and returns the average as a heatmap.

We evaluate the following classical XAI techniques: DeepSHAP, GradientSHAP~\cite{ge2022explaining, ge2022explainable}, and AttnLRP~\cite{achtibat2024attnlrp}. We apply RawWrapper, as described in Section \ref{sec:rawwrapper}, to obtain attributions in the time-frequency domain for the ADD model trained with raw waveform as input.  For DeepSHAP, we follow \cite{ge2022explaining, grinberg2025does} and use 20 random bona fide utterances for the reference value. The setup for GradientSHAP is identical to \cite{ge2022explainable, grinberg2025does} and generates 20 examples per utterances using zero-vector baselines. AttnLRP requires defining $\gamma$ hyperparameter for convolutional and linear layers outside of transformer. We do hyperparameter search with $\gamma \in \{0, 0.1, 10\}$ and choose the one that is more aligned with the dataset annotations (see Section \ref{sec:evaluation_criteria}): $10$ for 1d-convolution and $0.1$ for other layers.

\subsection{Results and discussion}

\begin{figure*}[!t]
    \centering
    \includegraphics[width=\linewidth]{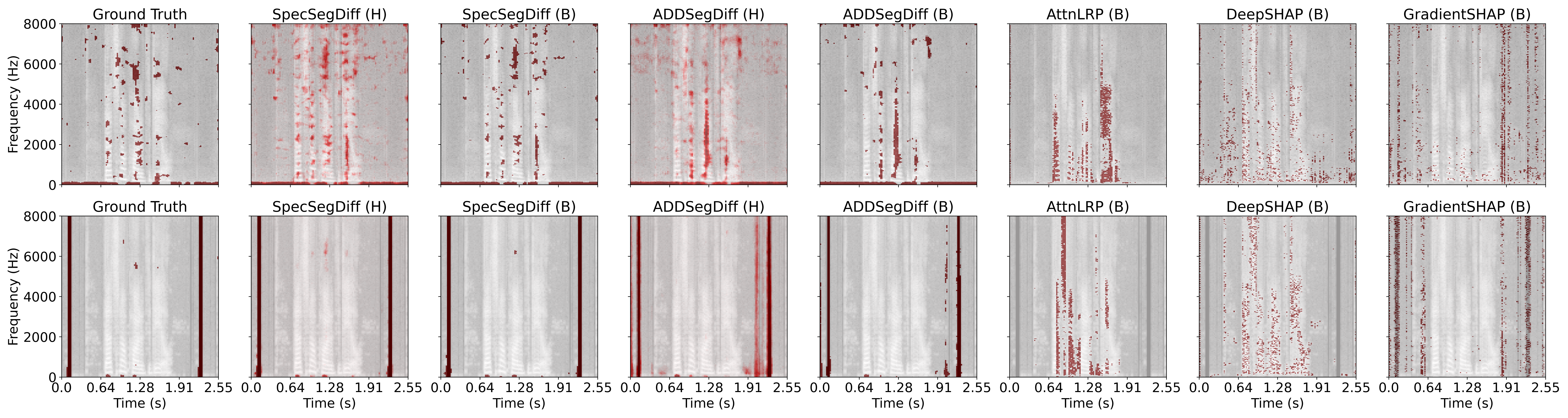}
    \caption{Comparison of the attributions from classic XAI tools and our diffusion model. We show binarized (B) and heatmap (H) versions. The binarization uses $95\%$ quantile threshold.}
    \label{fig:xai_comparison}
\end{figure*}

\begin{table*}[!t]
    \centering
    \caption{Metric for different XAI methods computed on the development set of VocV4 dataset (the first number) and test set of LibriSeVoc (the second number). Diffusion models evaluated on VocV4 and LibriSeVoc are trained on VocV4 and LibriSeVoc, respectively.}
    \label{tab:metrics}
    \resizebox{\linewidth}{!}{
    \begin{tabular}{l|ccccc|cccc}
    \toprule
    \multirow{2}{*}{Method} &  \multicolumn{5}{c|}{Segmentation} & \multicolumn{4}{c}{Faithfulness} \\
    & GDice $\uparrow$  & $F_1$ $\uparrow$  & IoU $\uparrow$  & FBound $\uparrow$  & SSIM $\uparrow$  & AI $\uparrow$ & AD $\downarrow$ & AG $\uparrow$ &  Fin-In $\uparrow$ \\
    \midrule
    DeepSHAP & $10.93$ / $8.78$ & $6.50$ / $4.24$  & $3.43$ / $2.21$  & $33.30$ / $19.81$ & $37.03$ / $33.29$ & $0.43$ / $1.82$  & $95.35$ / $92.74$  & $0.37$ / $1.62$  & $0.04$ / $0.03$ \\
    GradientSHAP & $11.71$ / $10.28$ & $7.31$ / $5.82$  & $3.99$ / $3.03$  & $30.45$ / $27.12$ & $33.55$ / $31.12$ & $0.53$ / $1.89$  & $91.03$ / $81.22$  & $0.38$ / $1.48$  & $0.08$ / $0.12$ \\
    AttnLRP & $12.65$ / $8.69$ & $8.30$ / $4.15$  & $4.49$ / $2.19$  & $30.29$ / $15.71$ & $53.22$ / $50.48$ & $0.42$ / $1.82$  & $96.38$ / $92.06$  & $0.37$ / $1.62$  & $0.03$ / $0.03$ \\
    \midrule
    ADDSegDiff & $48.26$ / $52.03$ & $45.69$ / $49.65$  & $31.23$ / $36.33$  & $57.36$ / $56.78$ & $69.04$ / $74.85$ & $0.81$ / $1.82$  & $68.34$ / $85.06$  & $0.50$ / $1.60$  & $0.31$ / $0.10$ \\
    \midrule
    SpecSegDiff & $57.46$ / $51.73$ & $55.34$ / $49.33$ & $40.51$ / $36.12$ & $66.95$ / $59.08$ & $70.99$ / $73.86$ & $-$ & $-$  & $-$  & $-$  \\
    \bottomrule
    \end{tabular}
    }
\end{table*}

Figure \ref{fig:xai_comparison} shows predictions from all the XAI methods on two vocoded audio from the VocV4 dataset. The utterances are re-synthesized from the same real audio but with different vocoders, Hn-NSF~\cite{wang2019neural} in the top row and WaveGlow~\cite{prenger2019waveglow} in the bottom row. The ground truth column shows that the artifacts introduced by these vocoders appear very distinct from each other. The first vocoder introduces artifacts in the low frequency bands across most of the timesteps as well as specific mid-frequency bands at particular timesteps corresponding to speech. However, the artifacts from the second vocoder appear to be independent of frequency bands and occur in non-speech regions at the start of audio. If we look at the predictions from classical XAI methods, we find that all the three methods fail to capture any detail in the ground truth for both utterances. Furthermore, their predictions for both audio are similar. Since both utterances are derived from the same real audio, we believe that these methods are simply highlighting characteristics of audio that, for example, may be related to content and not related to deepfake artifacts. On the other hand, both SpecSegDiff and ADDSegDiff have different outputs for the two utterances. They also seem to be better aligned with the ground truth, which is not surprising, since the diffusion models were trained for this purpose. In addition, SpecSegDiff predictions are the most accurate as it correctly highlights the lower frequency band and has a lot of points in approximately the same frequency bands for the middle part of the first audio. For the second audio, its predictions exactly match the ground truth. We note that since the frequency bands are arranged contiguously and multiple frequencies are grouped in a single band, it should not be a problem if predictions match approximately with the ground truth. For ADDSegDiff, we find a similar pattern. The predictions are much better than the classical XAI methods and match the ground truth, however, less accurately than SpecSegDiff.

We now compare the methods across entire dataset using quantitative metrics for the segmentation task, as well as faithfulness metrics. The results are shown in Table \ref{tab:metrics} on both VocV4 and LibriSeVoc datasets (separated by ``/''). We do not report faithfulness metrics for SpecSegDiff since faithfulness metrics are computed for ADD model, whereas SpecSegDiff does not rely on any ADD model. Next, we make 3 observations based on the table. Firstly, the numbers do not vary significantly across the two datasets for all the models and metrics, especially the segmentation metrics that measure the accuracy, as well as structural similarity between the prediction and ground truth. Secondly, there is a large gap between the metrics for classical XAI methods and diffusion-based methods. Finally, there is a smaller gap between SpecSegDiff and ADDSegDiff. The last two observations about the superior performance of diffusion-based methods match with our previous discussions based on the visual analysis of the predicted heatmaps. Moreover, the first observation suggests that our proposed method can be trained on any detection dataset that contains vocoder artifacts. Therefore, we believe that the proposed method leads to a superior XAI approach that is general enough to apply to any detection model and better aligned with ground truth explanations compared to existing techniques. We also note that the ground truth annotated regions are indeed used for the ADD model decision, since ADDSegDiff is able to overcome other methods in terms of faithfulness metrics and predict regions with high accuracy, while being conditioned only on the intermediate features.

Next, we address the performance gap between SpecSegDiff and ADDSegDiff models based on the above results in Figure \ref{fig:xai_comparison} and Table \ref{tab:metrics}.
Figure \ref{fig:xai_comparison} shows that ADDSegDiff misses a dense region (approximately in the middle of the ground truth) in the top row and predicts an extra column as shown in the continuous version of the heatmap in the bottom row. But the heatmaps predicted by SpecSegDiff are better and it also performs at least as good as ADDSegDiff on segmentation metrics. We believe that this could be due to the complex spatial relationship that ADDSegDiff needs to learn between latent space of the ADD model and time-frequency representation of the audio. The problem can be exacerbated due to the use of pretrained backbones in the ADD model. These backbones, such as Wav2Vec2~\cite{tak2022automatic}, are trained as an encoder-only model. So, there is no objective that encourages the model to preserve all the spatial information. However, the mapping is straightforward in the case of SpecSegDiff as it operates on a magnitude spectrogram directly. We hope that this problem can be resolved to some extent as we scale the training of the diffusion model to a larger dataset. We leave this as a future work.

\section{Conclusion}\label{sec:conclusion}
In this paper, we proposed a data-driven approach to generate explanations for the ADD task. We presented a methodology to create a supervised dataset containing spoof audio and corresponding ground truth explanation in the time-frequency domain. The dataset enabled a fair comparison of existing classical XAI techniques to generate the deepfake explanations. We used RawWrapper technique to extend the methods to the time-frequency domain. However, we found that these XAI tools are not able to handle the specificity required for the task. 
On the other hand, our diffusion-based XAI provides heatmaps that are more aligned with the ground-truth annotations and more faithful to the classifier in comparison to classic tools. 
As with any deep learning technique, model generalization to unseen vocoders, speakers, and other conditions is an important aspect. We believe that leveraging large pretrained diffusion backbones and scaling the training to diverse datasets containing annotations for many spoofing algorithms can improve robustness and be a potential future direction.

\bibliographystyle{IEEEtran}
\bibliography{main}

\end{document}